\documentclass{custom}
\usepackage{graphicx}
\usepackage{wrapfig}
\usepackage[section]{placeins} 
\usepackage{rotating}
\usepackage{amsmath}
\usepackage{amssymb}
\usepackage{mathptmx}
\usepackage{mathtools}
\usepackage{xfrac}
\usepackage[cal=cm]{mathalfa}
\usepackage[numbers,square]{natbib}
\usepackage{textgreek}
\usepackage{xcolor}

\makeatletter

\newcommand{\Ricochet}{\textsc{Ricochet}}

\newcommand{\one}{\textsuperscript{a}}
\newcommand{\two}{\textsuperscript{b}}
\newcommand{\fa}{\textsuperscript{$\star$}}

\begin{document}

\title{Optimization and performance of the CryoCube detector for the future \Ricochet{} low-energy neutrino experiment}

\author{%
	\name{T.~Salagnac\one\fa, J.~Billard\one, J.~Colas\one, D.~Chaize\one, M.~De~Jesus\one, L.~Dumoulin\two, J.-B.~Filippini\one, J.~Gascon\one, A.~Juillard\one, H.~Lattaud\one, S.~Marnieros\two, D.~Misiak\one, C.~Oriol\two, L.~Vagneron\one, for the \Ricochet{}~collaboration%
		\thanks{\fa{}\,Email: salagnac@ipnl.in2p3.fr}%
	}%
	\affil{%
		\one{}\,Univ. Lyon, Univ. Lyon 1, CNRS/IN2P3, IP2I, F-69622, Villeurbanne, France \\%
		\two{}\,IJCLab, CNRS, Univ. Paris-Sud, Univ. Paris-Saclay, 91120 Palaiseau, France%
	}%
}

\maketitle


\begin{abstract}

	The \Ricochet{} reactor neutrino observatory is planned to be installed at Institut Laue-Langevin starting in mid-2022. The scientific goal of the \Ricochet{} collaboration is to perform a low-energy and percentage-precision CENNS measurement in order to explore exotic physics scenarios beyond the standard model. To that end, \Ricochet{} will host two cryogenic detector arrays : the CryoCube (Ge target) and the Q-ARRAY (Zn target), both with unprecedented sensitivity to O(10) eV nuclear recoils. The CryoCube will be composed of 27 Ge crystals of 38g instrumented with NTD-Ge thermal sensor as well as aluminum electrodes operated at 10\,mK in order to measure both the ionization and the heat energies arising from a particle interaction. To be a competitive CENNS detector, the CryoCube array is designed with the following specifications : a low energy threshold ($\sim 50$\,eV), the ability to identify and reject with a high efficiency the overwhelming electromagnetic backgrounds (gamma, betas, X-rays) and a sufficient payload ($\sim 1$\,kg).
	After a brief introduction of the future Ricochet experiment and its CryoCube, the current works and first performance results on the optimization of the heat channel and the electrode designs will be presented. We conclude with a preliminary estimation of the CryoCube sensitivity to the CENNS signal within \Ricochet{}.


\end{abstract}


\section{Introduction}


\subsection{The \Ricochet\ experiment}

The recent first observation of Coherent Elastic Neutrino-Nucleus Scattering (CENNS) by the COHERENT collaboration has opened new avenues to search for new physics. These include the existence of sterile neutrinos and of new mediators as well as Non Standard Interactions~\cite{Billard_2018}. The \Ricochet{} collaboration~\cite{JLTP_Ricochet} is aiming for a percentage-level CENNS measurement down to sub-100 eV recoil energies where signatures of such new physics may arise. The future \Ricochet{} experiment will take place at the Institut Laue-Langevin where low-threshold cryogenic detector arrays, the Q-Array and the CryoCube, will be deployed at about 8.8 meters away from a 58 MW research reactor in 2022/2023.

From the proximity of the reactor core and being at the surface level with only 15\,m.w.e of overburden, \Ricochet{} will operate in a high background environment. To achieve a manageable background level of less than 100 evt/kg/keV/day an heavy shielding will be deployed around and inside the cryostat. The latter is composed from the closest to the farthest layer from the detectors of: 35\,cm of borated polyethylene followed by 20\,cm of lead and a muon-veto with almost 4\textpi{} coverage. Finally, to reach the goal of a signal-to-background ratio greater than 1, particle identification is a key feature of \Ricochet{} detectors to separate electronic recoils (ER), produced by electromagnetic background like gammas, from nuclear recoils (NR) produced by neutrinos and neutrons.

With such specifications, the \Ricochet{} experiment is expected to reach a 5-sigma level CENNS detection significance in a couple of days and will be able to put strong constrains on new physics scenarios after one year of data taking.


\subsection{The CryoCube detector array}

\begin{wrapfigure}{O}{0.43\linewidth}
	\vspace{-20px}
	\begin{center}
		\includegraphics[height=4.5cm, keepaspectratio]{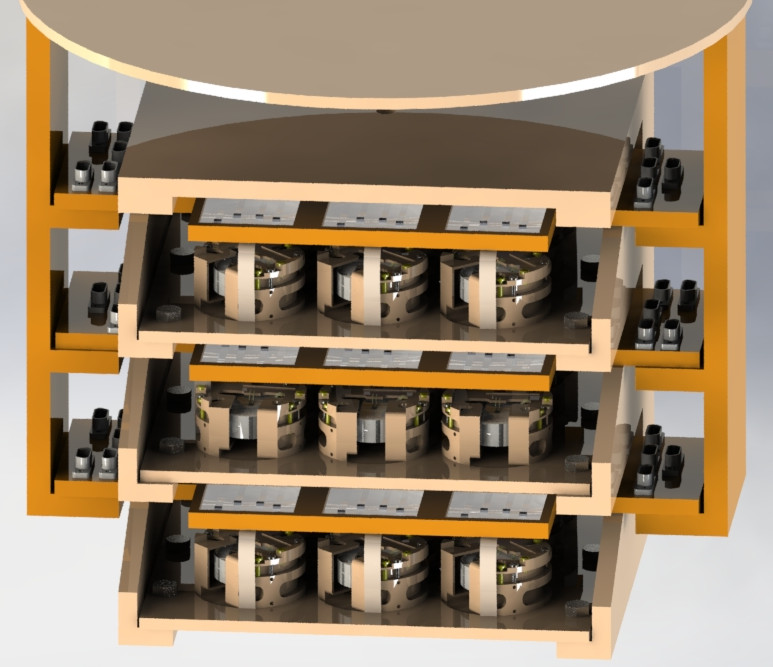}
	\end{center}
	\caption{Draft model of the CryoCube design. All the parts in dark orange correspond to 1\,K stage holding the electronics. The light part supporting the 27 germanium detectors  will be regulated at about 15\,mK. For the sake of visibility, the copper tight-infrared boxes for each individual detector are not represented.}
	\label{fig:CryoCube}
	\vspace{-10px}
\end{wrapfigure}

The CryoCube will consist of an array of 27 ($3\times3\times3$) high purity germanium crystal detectors, encapsulated in a radio-pure infrared-tight copper box suspended below the inner shielding. Each detector mass is about 38\,g to reach a total target mass around one kilogram. A $\mathcal{O}(10)$\,eV energy threshold is desired as the discovery potential scales exponentially with lowering the energy threshold. Considering a 50\,eV energy threshold, about 12.8\,evts/day of CENNS interactions are expected in the CryoCube detector array.
To reach such threshold, Ricochet germanium detector will be equipped with germanium neutron transmutation doped sensors (NTD) allowing to sense the heat energy from recoils. To achieve particle identification, the detectors will have a double heat/ionization readout. Ionization measurement is realized thanks to aluminum electrodes allowing to apply an electric field and collect signals from the ionization electron drift across the crystal.

In order to efficiently reject a significant fraction of the cosmogenic background, a muon-veto will be used in anti-coincidence with the target cryogenic detectors. Due to their $\mathcal{O}(100)$\,Hz muon-induced trigger rate, fast enough detectors with at least a $\mathcal{O}(100)$\,\textmu{}s timing resolution are required to minimize the muon-veto induced livetime loss~\cite{hdr_julien}.


\section{Heat channel optimization performance}


\subsection{Heat resolution optimization and performance}

\begin{figure}[htbp]
	\begin{minipage}{0.48\textwidth}
		\centering%
		\includegraphics[height=4cm, keepaspectratio]{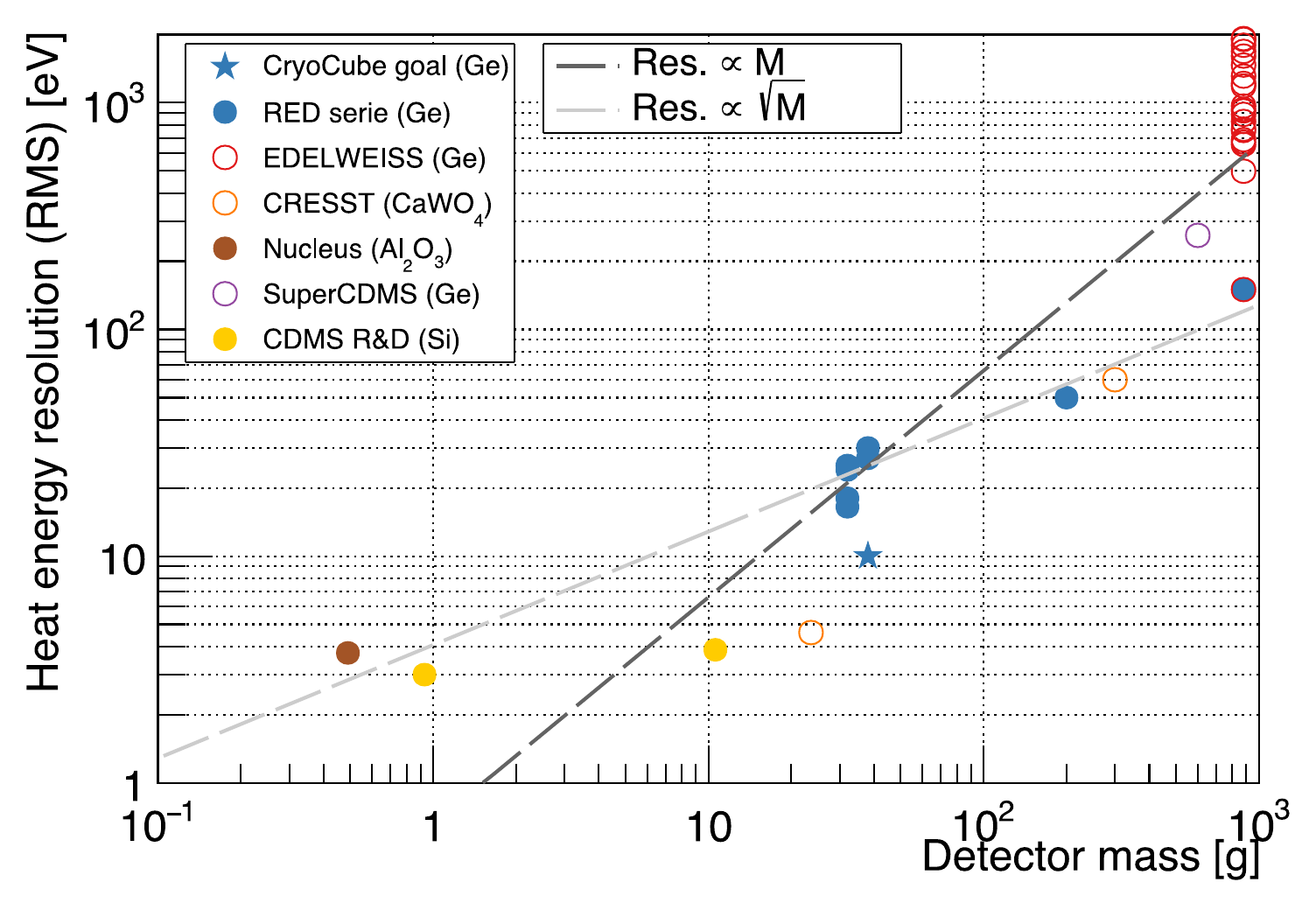}
	\end{minipage}%
	\hfill%
	\begin{minipage}{0.48\textwidth}
		\centering%
		\includegraphics[height=4cm, keepaspectratio]{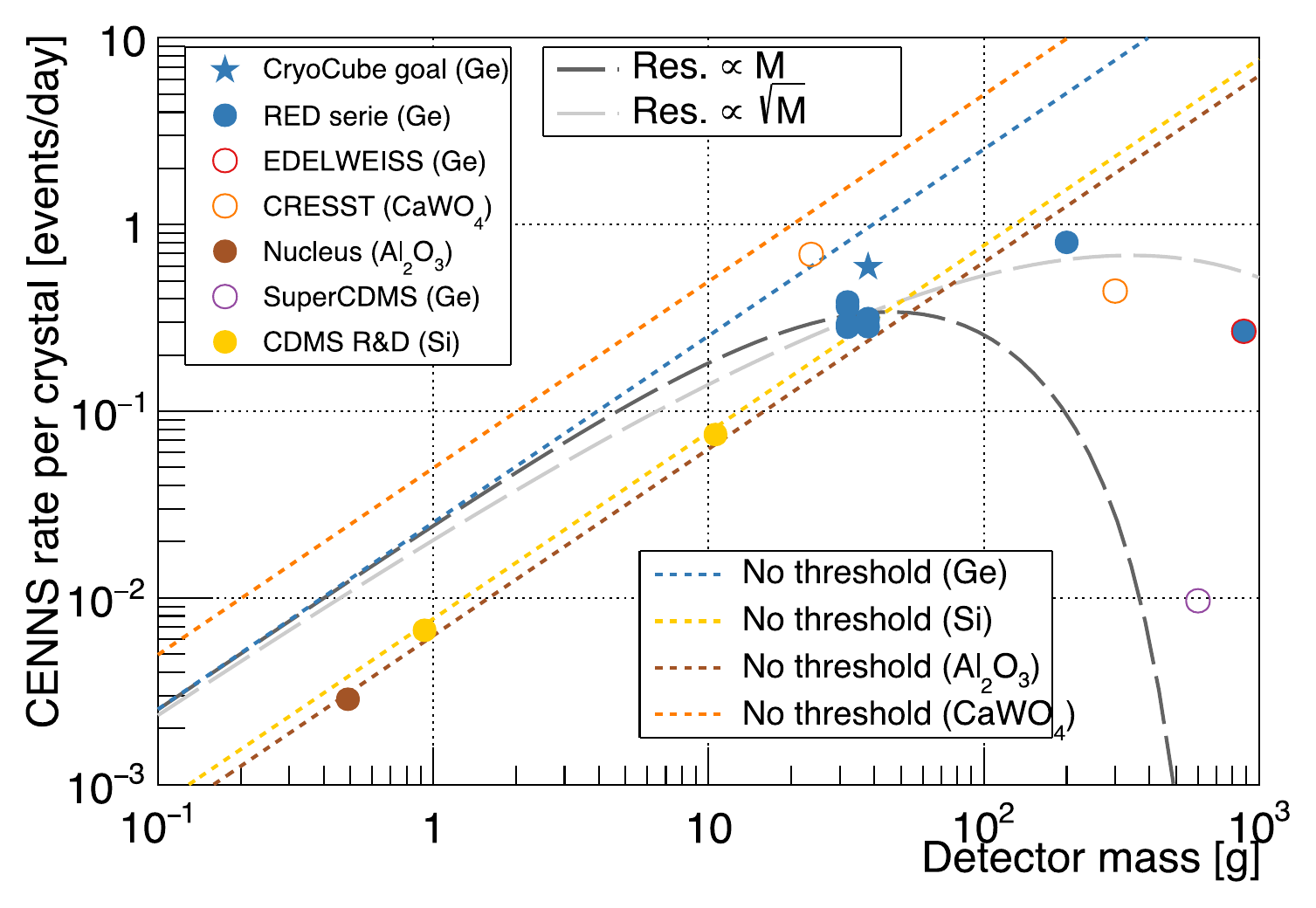}
	\end{minipage}
	\caption{\textit{Left:} The filled and empty circles correspond respectively to above- and under-ground measurements. The dashed dark and light gray lines represent two different scaling laws of the energy resolution with the detector mass. \textit{Right:} The CENNS rate has been computed considering a reactor neutrino flux of $10^{12}$ neutrinos/cm$^2$/s and a common threshold definition of 5\,\textsigma$_E$. The thin dotted lines correspond to the expected CENNS event rate considering a 0\,eV energy threshold for the relevant target materials.}
	\label{fig:mass_optimization}
\end{figure}

A comprehensive electro-thermal model has been developed~\cite{These_Dimitri} to optimize the heat energy resolution of the CryoCube germanium detectors which are equipped with a single NTD heat sensor. The individual detector mass has been first validated with 33.4\,g germanium crystal prototypes with an average heat channel baseline resolution of 22\,eV (RMS) on five detectors. The best resolution achieved, 17\,eV (RMS), was obtained with AC JFET-based EDELWEISS electronics at surface level~\cite{RED20_surf}. This resolution is 30\% higher than our thermal model predictions considering the JFET-based electronic noise with a 400\,Hz modulation frequency (13.2\,eV). This is mostly explained by some variance in the NTD sensitivity which depends on the intrinsic NTD characteristics ({\it e.g.} precise doping level).

Figure~\ref{fig:mass_optimization} provides a state of the art of various cryogenic detector performance as a function of their masses. This figure suggests that thanks to their 30\,g-scale mass and 20\,eV (RMS) resolution, the individual detector prototypes (blue circles) have proven to be expecting the highest CENNS event rate of 0.3\,events/day (0.6\,events/day as per the CryoCube expected 10\,eV resolution) from surface operations. According to our theoretical scaling laws, one can see that we could have envisioned higher masses, up to a $\sim 300$\,g crystals to improve by a factor 2.5 the expected CENNS rate per crystal. However, our goal is not only to have the largest CENNS rate, but also the highest sensitivity to many new physics signals that may predominantly arise at the lower end of the CENNS induced recoil energy spectrum. Such sensitivity to new physics is illustrated by the distance between the dots and the corresponding dashed line (0~eV energy threshold)~\cite{hdr_julien}. Also, larger crystals usually induce slower time response which may affect the muon-veto tagging efficiency which is pivotal in the context of above-ground operation as required for CENNS searches.

Lastly, to achieve the CryoCube specifications, we are developing low-noise  HEMT-based preamplifiers which, combined with our thermal model predictions, should lead to a baseline heat energy resolution of 10~eV (RMS)~\cite{Juillard2020, JLTP_HEMT}, hence improving by about 40\% the resolution with respect to our currently used JFET-based AC-modulated electronics.


\subsection{New holder design}

\begin{wrapfigure}{O}{0.5\linewidth}
	\vspace{-20px}
	\begin{center}
		\includegraphics[height=4.5cm, keepaspectratio]{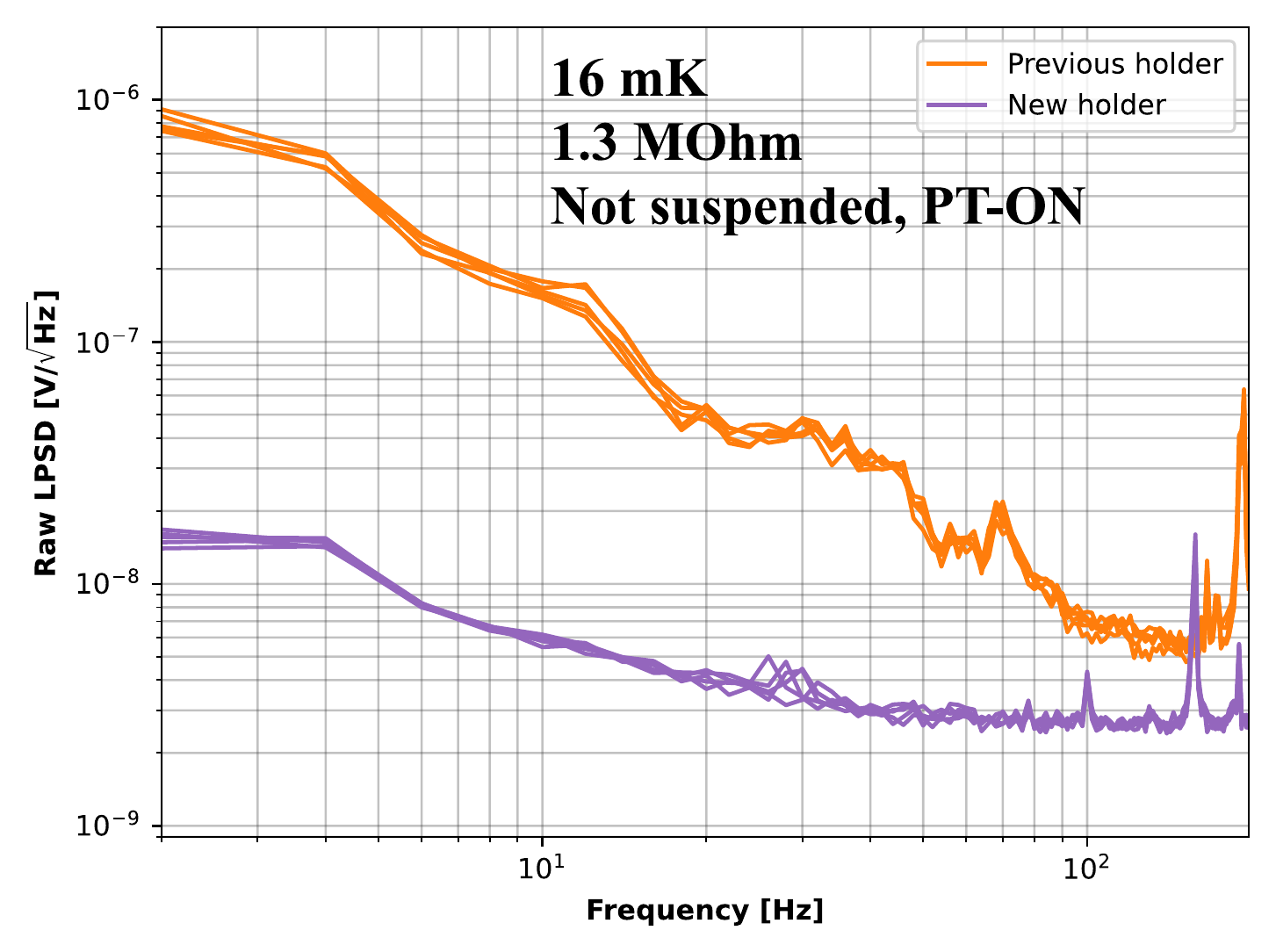}
	\end{center}
	\caption{Heat channel power spectral density measured for the same crystal and NTD but with the previous and new holders, at 16\,mK, with a NTD resistance of 1.3 M\textOmega{}, without any cold suspension and with the cryostat pulse-tube ON.}
	\label{fig:holder_psd}
	\vspace{-15px}
\end{wrapfigure}

A new detector holder has been designed to improve microphonic noise. The crystal in the previous holder was maintained thanks to 3 Teflon clamp on the bottom and 3 sapphire balls held by chrysocale/bronze clamps on the top. For an improved control of the clamping force, in the new design the Teflon clamps have been replaced by sapphire balls. Additionally, to constrain lateral displacements of the crystal we added 3 sapphire balls on the side, also held by chrysocale/bronze clamps. The crystal movement is therefore constrained along all three axes with adjustable rigidity. Figure~\ref{fig:holder_psd} shows the heat channel power spectral density measurements for the previous and new holder designs with the same detector and NTD sensor, and in the same operating conditions with no cryogenic suspensions. We observed a reduction by 2 orders of magnitude of the vibration induced noise at 1 Hz. Similar results have been confirmed with 3 other detectors, hence validating this new detector holder design.

Further optimizations are ongoing, {\it e.g.} tune holding stress from the sapphire balls and minimizing kapton capacitance to only a few pF. Our goal with the holder design optimization is to reach the targeted performance without the need of cryogenic suspension~\cite{Maisonobe:SuspensionCryo}.


\section{Detector electrode design}


\subsection{Electrostatic simulations}

\begin{figure}[htbp]
	\begin{minipage}{0.48\linewidth}
		\centering
		\includegraphics[width=0.98\linewidth, keepaspectratio]{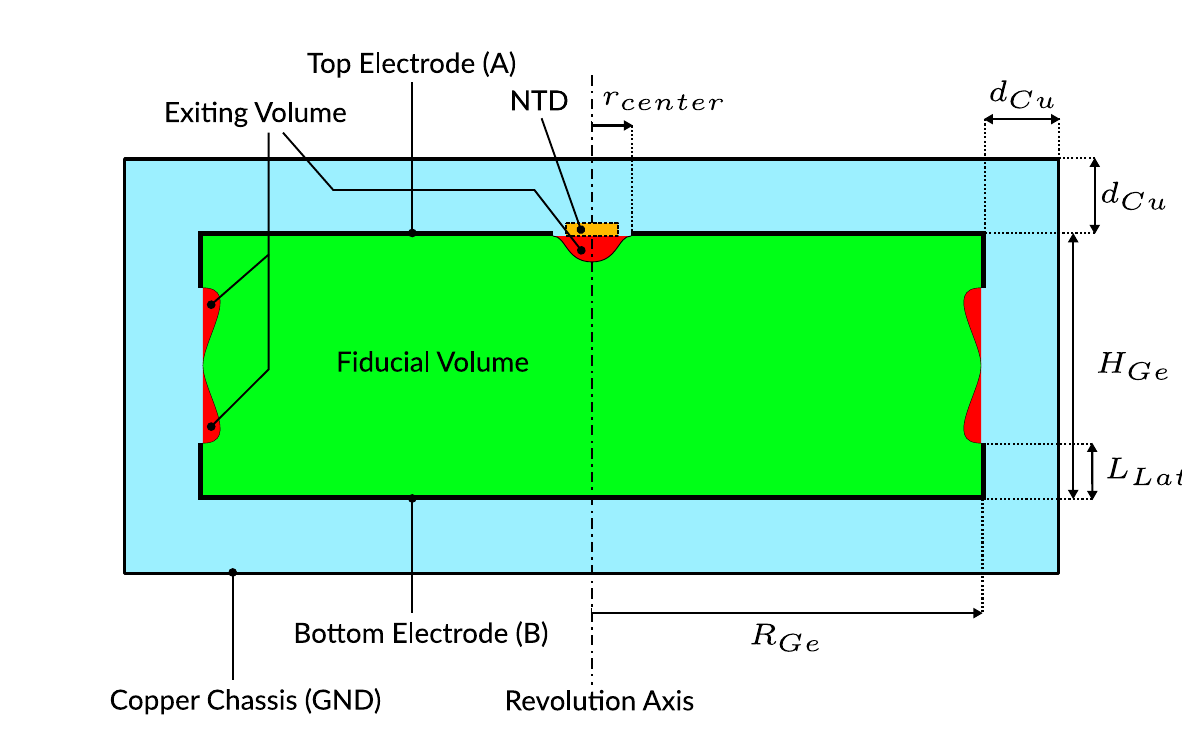}\\%
		\includegraphics[height=2.8cm, keepaspectratio]{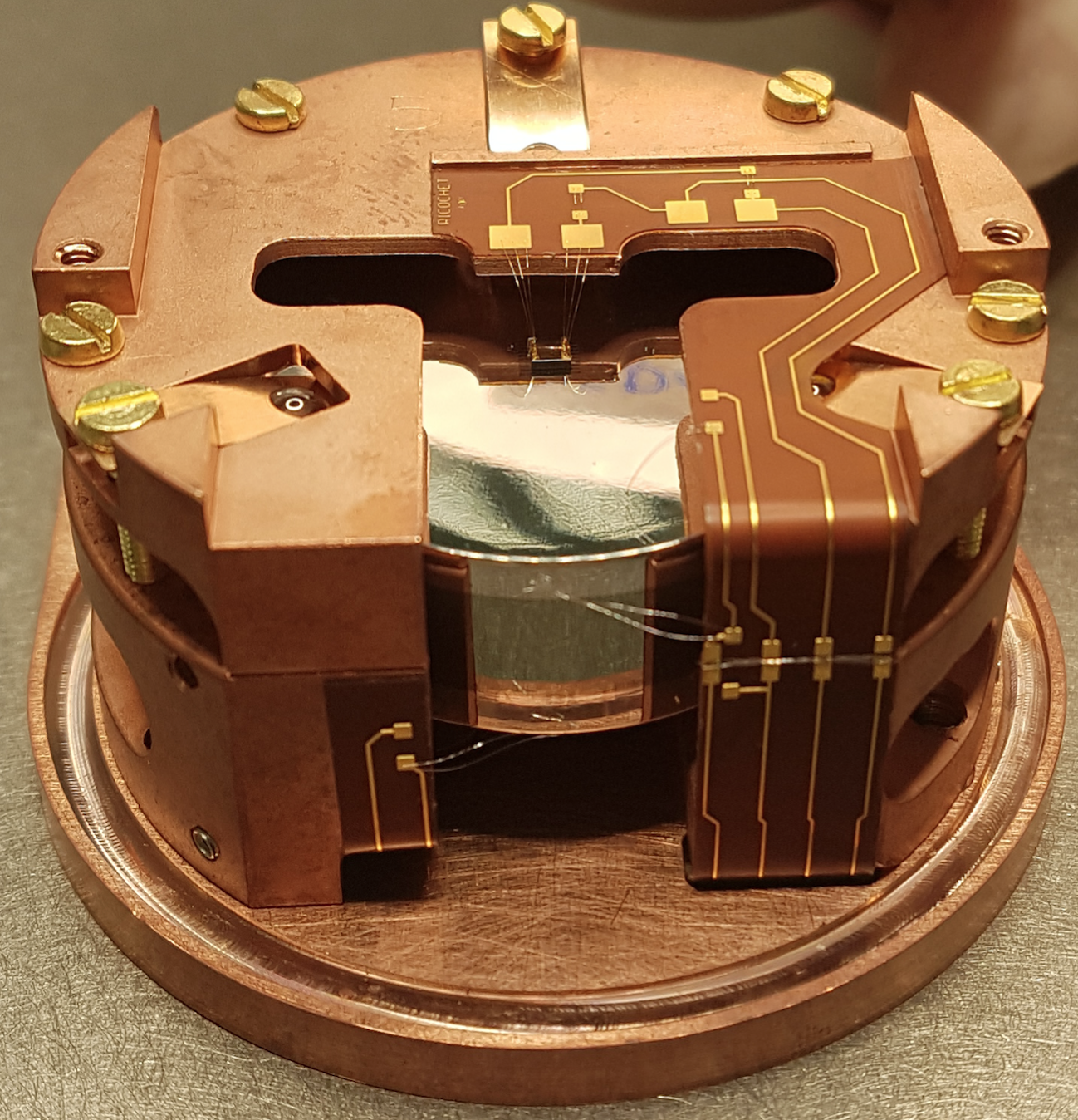}
	\end{minipage}
	\begin{minipage}{0.48\linewidth}
		\centering
		\includegraphics[width=0.98\linewidth, keepaspectratio]{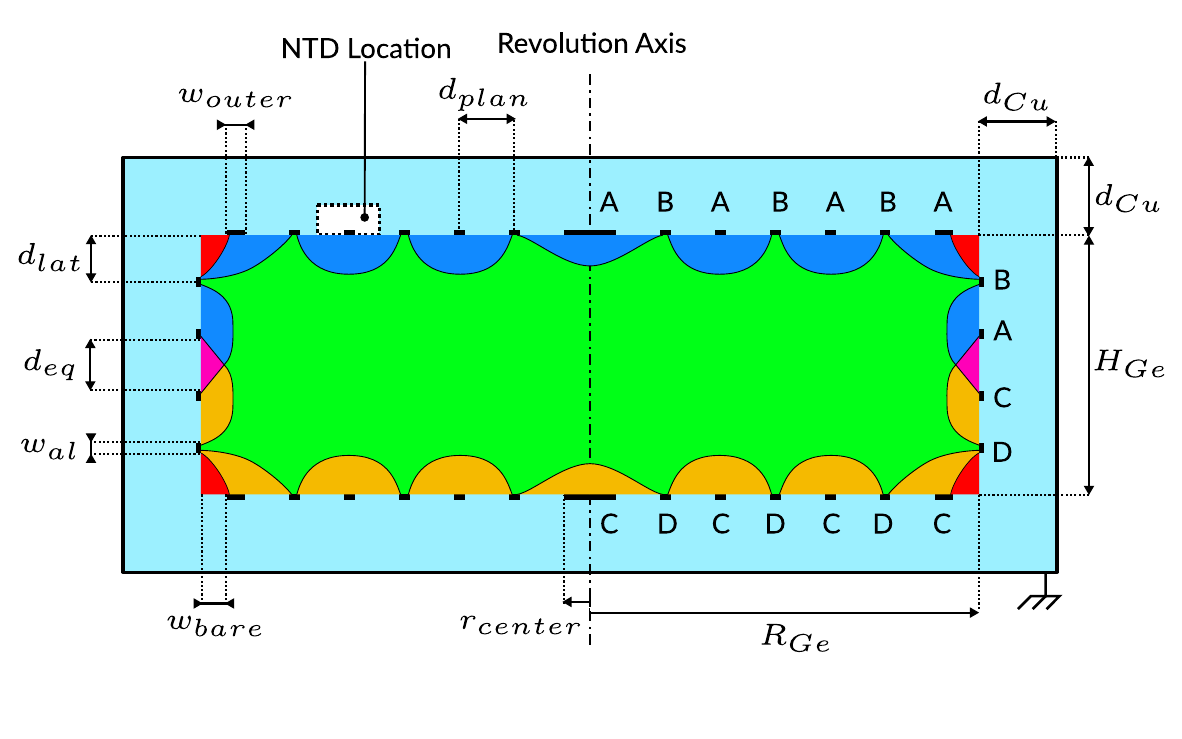}\\%
		\includegraphics[height=2.8cm, keepaspectratio]{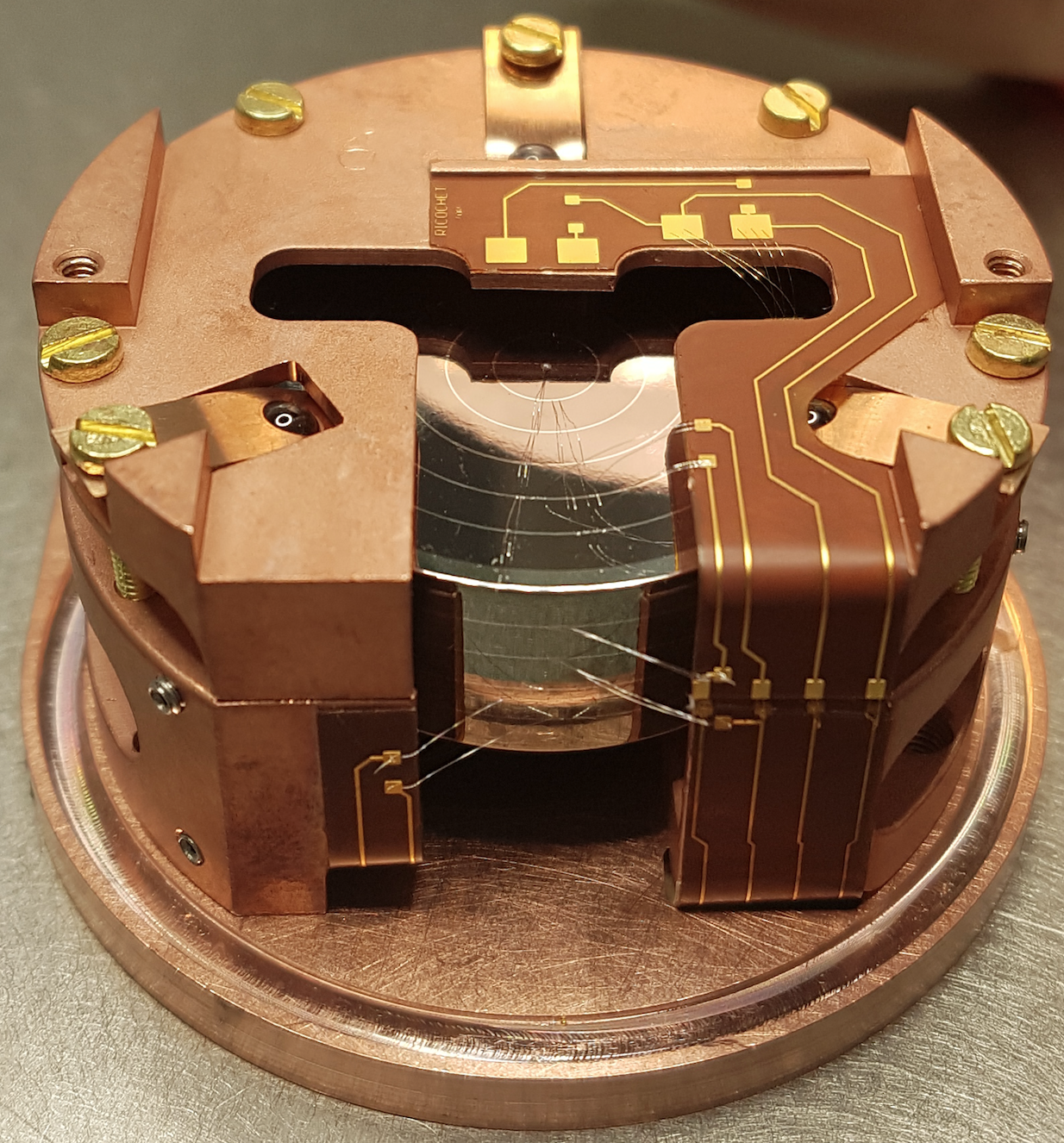}
	\end{minipage}
	\caption{Detailed PL (\textit{left}) and FID (\textit{right}) detector designs. Schematics (\textit{top}) taken from~\cite{These_Dimitri} exhibiting all relevant design parameters and charge collection zones: fiducial (\textit{green}), corner (\textit{red}), equatorial (\textit{pink}), top veto (\textit{blue}), and bottom veto (\textit{yellow}). Pictures of prototypes (\textit{bottom}), the PL and FID detectors are inside newly designed holders.}
	\label{fig:schema_detectors}
\end{figure}

Two electrode designs, presented in Fig.~\ref{fig:schema_detectors}, are being considered for the CryoCube: planar detector (PL), with one electrode on the top and another one on the bottom of the crystal, and fully interdigitated detector (FID), with ring electrodes covering the entire crystal surfaces. FID detector use four kinds of electrodes, two fiducial electrodes and two veto electrodes allowing to sense the ionization from  events occurring at the crystal surface (region in blue and yellow).
According to HEMT preamplifier noise model~\cite{Juillard2020}, an ionization baseline energy resolution of 20\,eVee, needed for particle discrimination down to 50\,eV threshold, could be reached with low capacitance electrode (below 20\,pF).

Both electrode designs have been optimized thanks to Multiphysics and AC/DC modules from COMSOL software. The PL and FID are characterized by 5 and 11 parameters respectively, hence resulting in computationally intensive high dimensional scans~\cite{These_Dimitri}.

According to our optimization, while FID design allows for surface event rejection strategy, it also comes with three drawbacks: 1) reduced fiducial volume (70\%), 2) low-electric field in the bulk ($E>0.2$\,V/cm) possibly degrading the charge collection efficiency, and 3) larger capacitance (18\,/\,16\,pF), due to the proximity of neighboring electrodes, which will result in degraded resolutions. The PL design should lead to an improved charge collection efficiency thanks to its higher bulk electric field ($E>0.9$\,V/cm) and have a much larger fiducial volume (99.2\%), but at the cost of not being able to reject surface events. The choice between these two designs will ultimately be done to maximize the Ricochet background mitigation, and both designs are therefore being investigated for the time being.


\subsection{Characterization of prototype designs}

\begin{figure}[htbp]
	\begin{minipage}{0.48\textwidth}
		\centering%
		\includegraphics[width=0.95\linewidth]{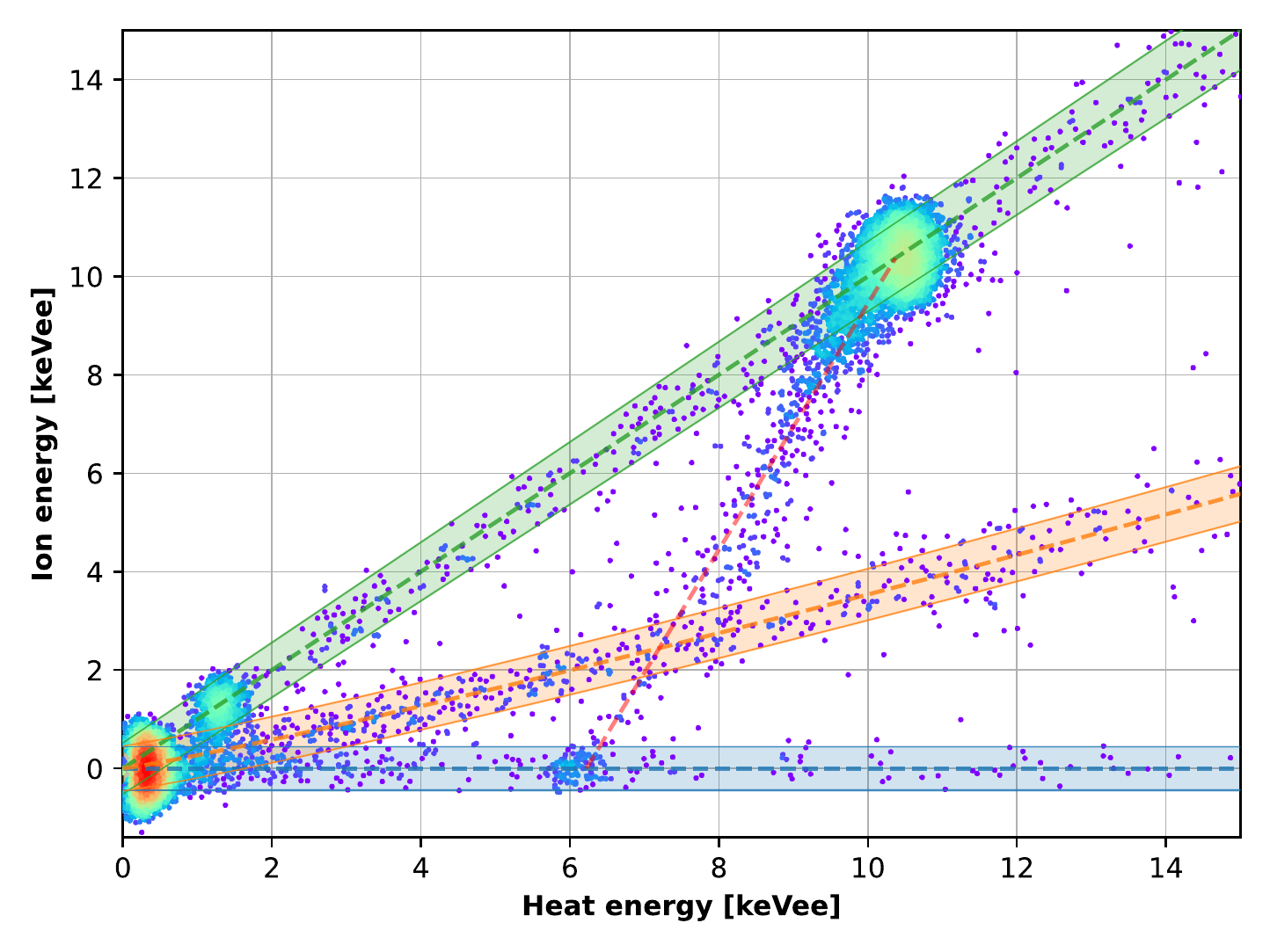}%
	\end{minipage}
	\hfill%
	\begin{minipage}{0.48\textwidth}
		\centering%
		\includegraphics[width=0.95\linewidth]{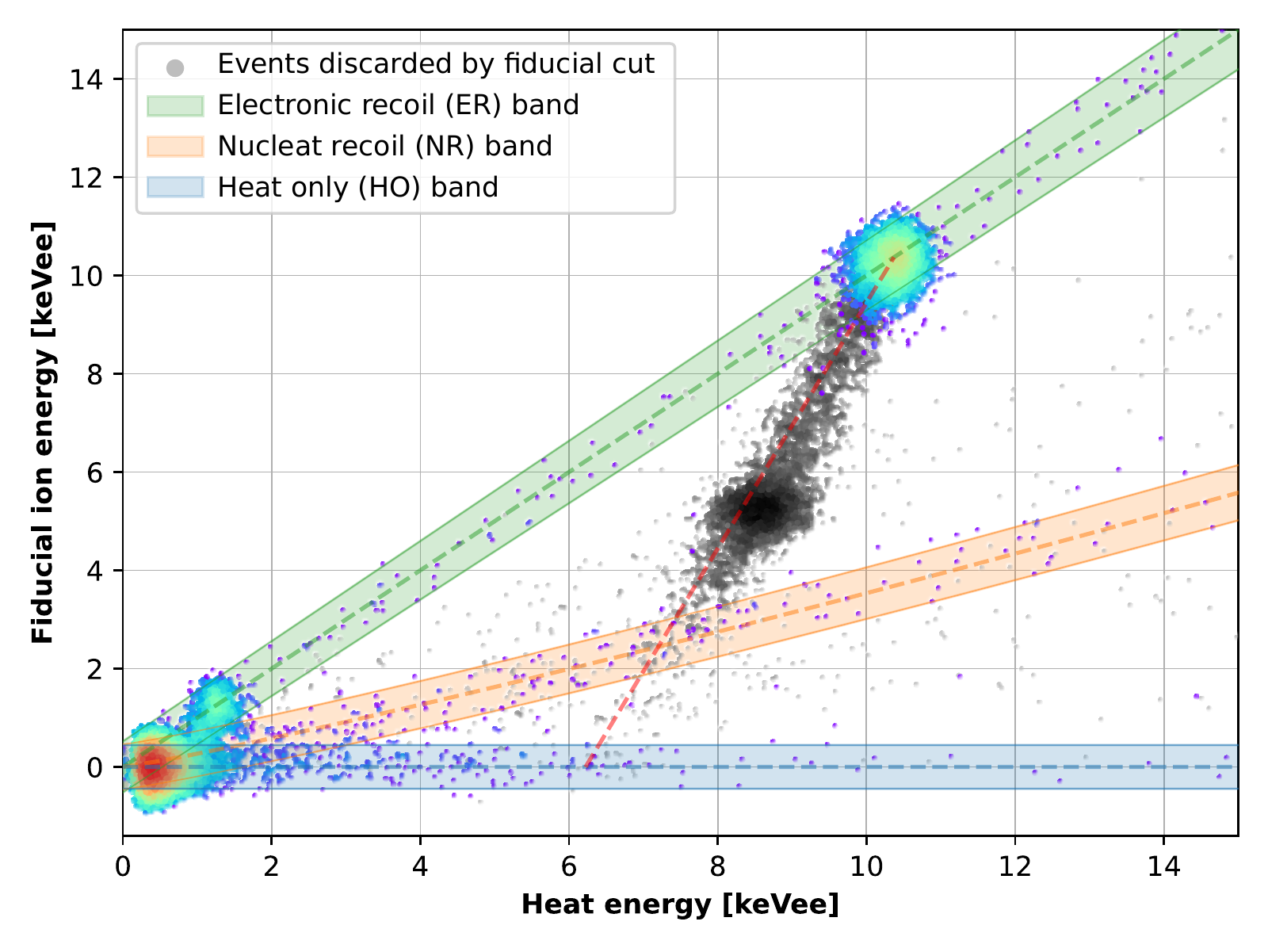}%
	\end{minipage}%
	\caption{Ionization energy on the collecting electrodes as a function of the heat energy given in electron equivalent (keVee) for about 48h of livetime with electrode bias voltage of 2V, for a PL prototype on the \textit{left} and an FID prototype on the \textit{right}.}
	\label{fig:design_charac}
\end{figure}

PL and FID detector prototypes have been operated above-ground inside a cryostat surrounded by a light 10-cm thick lead shielding at IP2I Lyon. About 2 days of data have been used to measure the detector performance with EDELWEISS JFET-based electronics, within the following experimental conditions: crystal temperatures regulated around 20\,mK and electrode voltage bias of 2\,V. Prior to their cooling, the germanium crystals were activated by a neutron source, allowing energy calibration from the K/L shell EC-decays of the $\prescript{71}{}{\mathrm{Ge}}$ isotope, respectively at 10.37\,keV and 1.3\,keV. We used the same processing pipeline as in~\cite{RED20_surf, JLTP_MPS}, based on  optimal filter event reconstruction. Detector performance have been evaluated on events passing quality cut selections ({\it e.g.} pulse fit $\chi^2$ and non-saturated offsets).

The resulting selected event distribution in the fiducial ionization versus heat energy plane is presented in Fig.~\ref{fig:design_charac}. Fiducial volumes, corresponding to the green area on figure~\ref{fig:schema_detectors} are estimated differently for the PL and FID designs. The PL fiducial volume is considered to be where the charge conservation condition is respected, {\it i.e. } sum of all the ionization channel amplitudes equals to zero. Considering 3-$\sigma$ band tolerance cuts, we found the PL fiducial volume to be above 98\%. For the FID prototype, charges from bulk events are collected only by the fiducial electrodes (B and D). The fiducial volume is then defined as the ratio of events with null energy on the veto electrodes (A and C) within a 2 sigma tolerance cut. By taking the ratio between the number of events passing the fiducial cut (colored events on right panel) and the total number of events, we found a fiducial volume around 62\% for the FID prototype. Due to charge trapping some events can have an incomplete charge collection, degrading particle identification capabilities since it changes the ratio of heat/ionization energies. Such events are clearly visible with the PL prototype (left panel) and are distributed according to the red dashed line from the 10.37\,keV peak to the heat only region. Fractions of incomplete charge events have been defined as the fraction of 10.37\,keV events outside the 2\,\textsigma{} peak region. Events were selected using handmade selections. By taking care of subtracting the different contribution of ER, NR and heat only events, we obtained superior limits on the fraction of incomplete charge collection events of 10\% for the PL and 1\% for the FID designs. Leakage of these events in the NR band, where we expect the CENNS signal, appears to be inferior to 1\% and 0.3\% respectively for the PL and the FID prototypes. Estimated performance are close to the ones predicted by the COMSOL simulations. In addition, surface event characterizations, using dedicated sources, for both designs are ongoing.

The final decision regarding the choice between these two detector designs will be done considering the observed ILL background at low energy with the upcoming HEMT preamplifier which are expected to lower the particle identification threshold from 1\,keV to 100\,eV.


\section{CryoCube CENNS projected sensitivity}

\begin{figure}[htbp]
	\begin{minipage}{0.48\textwidth}
		\centering%
		\includegraphics[width=0.95\linewidth, height=3.3cm]{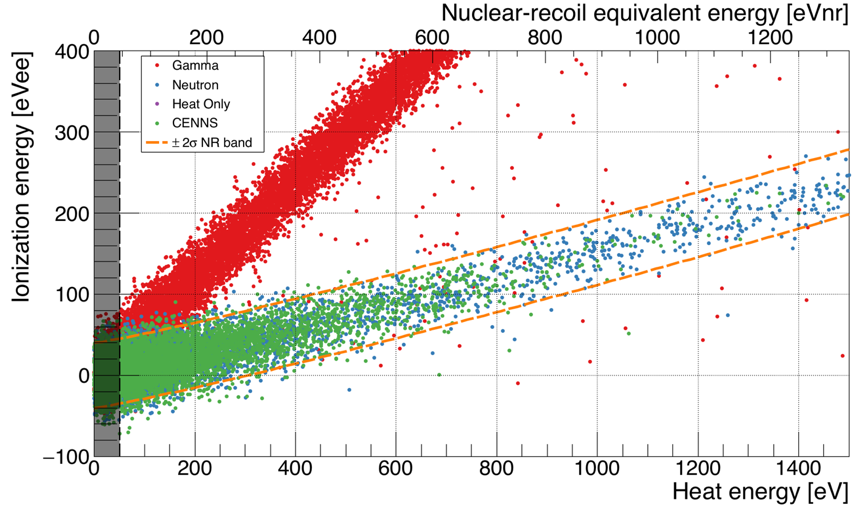}\\%
		\includegraphics[width=0.95\linewidth, height=3.3cm]{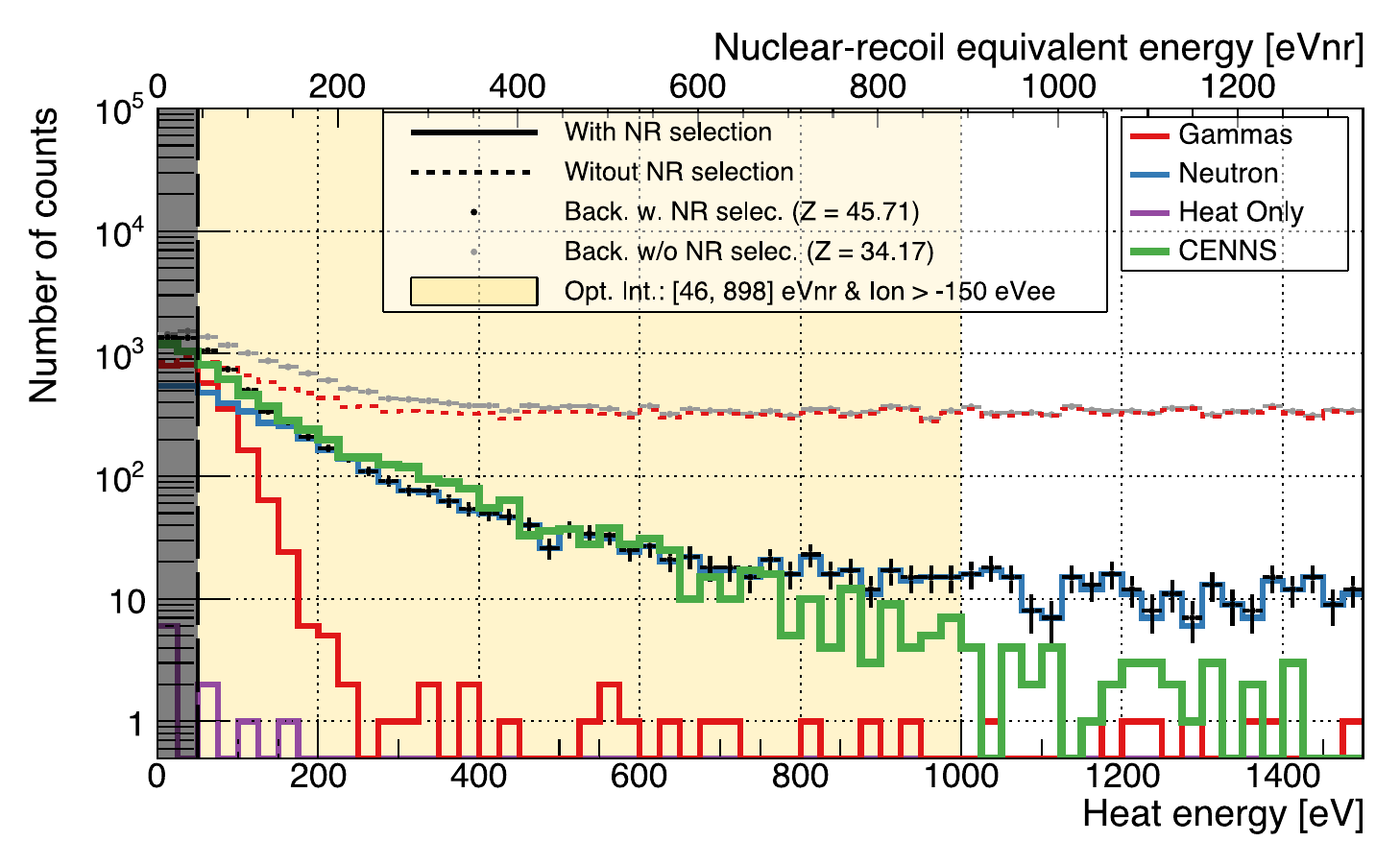}
	\end{minipage}%
	\hfill%
	\begin{minipage}{0.48\textwidth}
		\centering%
		\includegraphics[width=0.95\linewidth, height=3.3cm]{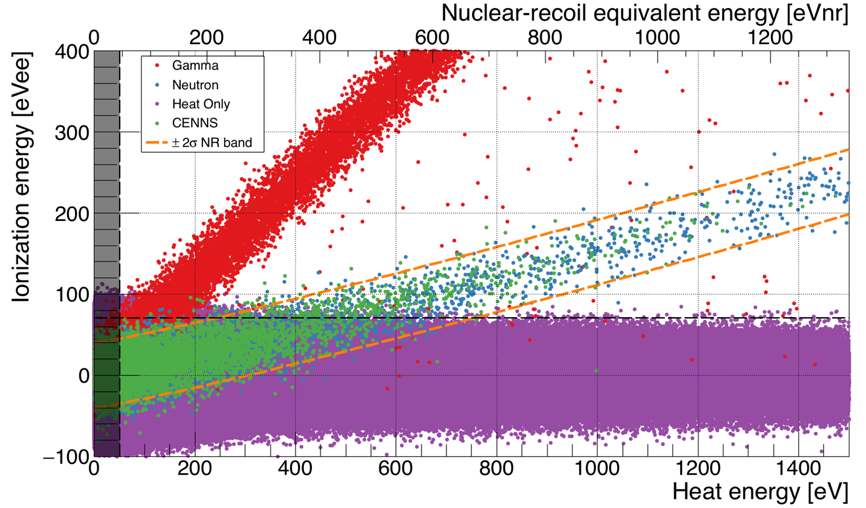}\\%
		\includegraphics[width=0.95\linewidth, height=3.3cm]{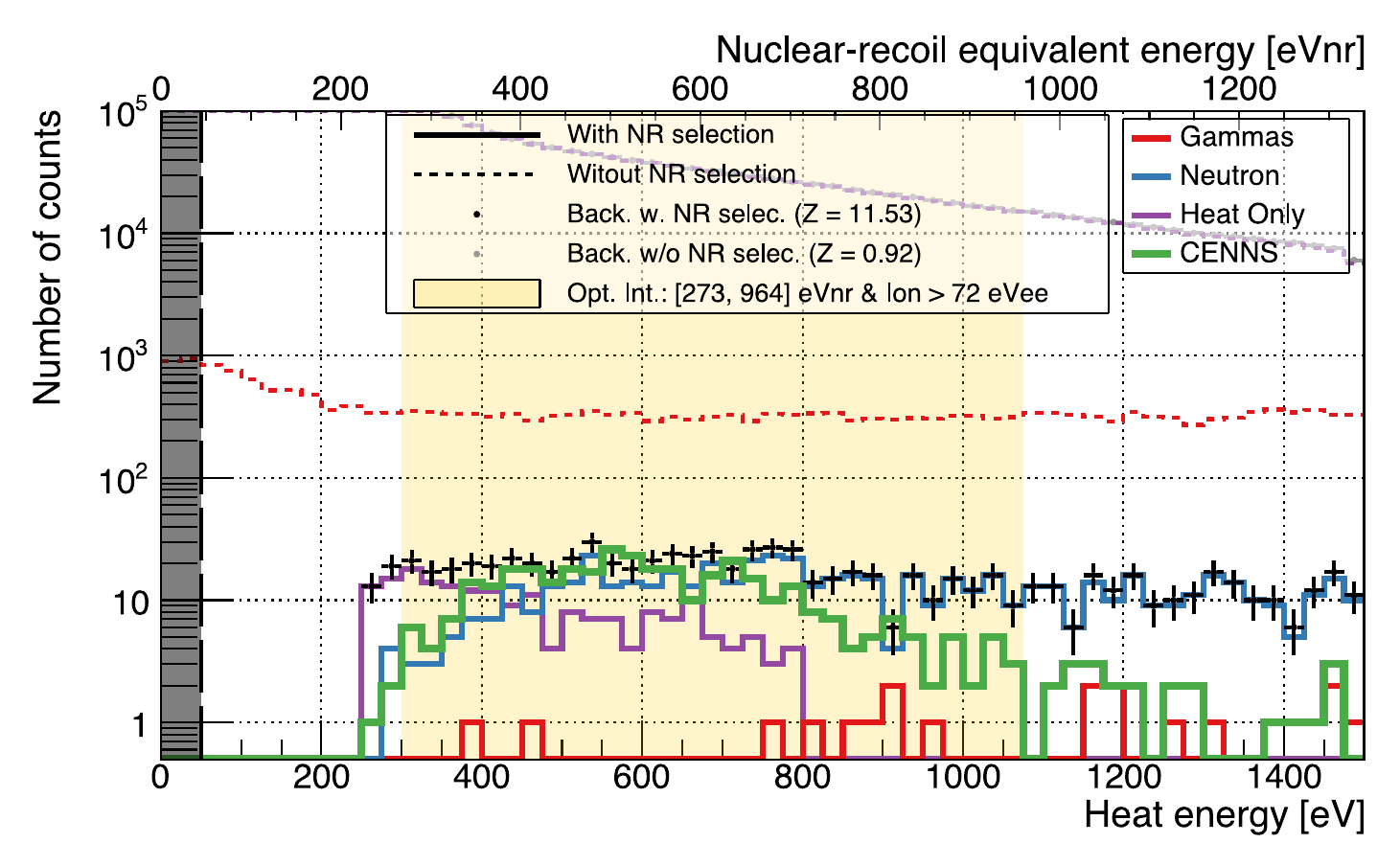}
	\end{minipage}
	\caption{\textbf{Top:} Simulated event distribution on the ionization vs heat plane where the CENNS signal, the gamma, the neutron and the hypothetical heat only (right panels) backgrounds are shown as green, red, blue and violet dots respectively. The nuclear recoil candidates are selected from the $\pm 2\sigma$ nuclear recoil (NR) band (orange dashed lines) and an optimized ionization threshold (horizontal black dashed line). \textbf{Bottom:} Corresponding heat energy distributions with (without) the NR selection as solid (dashed) lines or dots. The yellow area illustrates the optimal energy interval. In both panels, the black filled region illustrates the 50 eV heat energy threshold. We considered here the CryoCube specifications and the muon veto with $\Delta T_{coinc} = 350$\,\textmu{}s~\cite{hdr_julien}.}
	\label{fig:sensitivity}
\end{figure}

The sensitivity of the future CryoCube detector array to the observation of the CENNS signal has been estimated using a MC simulation and considering only statistical uncertainties. The simulation takes as input the expected energy spectra of CENNS and ER/NR from simulated ILL backgrounds. The ILL background model takes into account the reactogenic and cosmogenic neutron and gamma components of the total background, obtained from a detailed GEANT4 simulation of the Ricochet setup ant its shielding at ILL. Figure~\ref{fig:sensitivity} shows the resulting simulation for 350 days of reactor ON data at ILL, corresponding to about 7 cycles distributed over 2 years. A "worst case" scenario is presented on the right panel taking into account an additional low-energy background excess producing similar rate of heat only (HO) events -- events with null ionization yield -- as observed in previous detectors~\cite{edw-red30, JLTP_JulesG}.

The energy threshold in the baseline scenario is 50\,eV in accordance with the CryoCube specifications. In the worst case scenario, an effective CENNS energy threshold is increased to 250\,eV due to an ionization threshold mitigating the HO background. After only 1-cycle (50 days), the Ricochet experiment is expected to reach a CENNS detection significance of 17.3\,\textsigma{} and 4.3\,\textsigma{}, and a 2\%-to-8\% (stat. only) CENNS precision measurement after 2 years of reactor ON data, with or without the presence of this strong low-energy excess~\cite{hdr_julien}. Note that without particle identification a significant CENNS detection appears to be unreachable. This shows how crucial particle identification is in the context of this possible low-energy background to CENNS searches.


\section{Conclusion}

We have presented the modeling, optimization and first experimental characterizations of the CryoCube detector design candidates (PL and FID) to be integrated within the future \Ricochet{} experiment at ILL by 2022/2023.
Its key feature is to push for particle identification down to $\sim$100~eV to reject both known and unknown backgrounds.
With the targeted performance, a CENNS detection after only one reactor cycle (50 days) at the 4.2-to-17.3 sigma level is expected.
Such precision measurement after two years will lead to unprecedented sensitivity to various new physics scenario.


\section*{Acknowledgements}
We are grateful to C.~Nones and X.-F.~Navick for their help with detector fabrication and we thank the EDELWEISS collaboration as this work has been done within the synergetic R\&D programs between EDELWEISS-SubGeV and \Ricochet{}/CryoCube.
This work is supported by the European Research Council (ERC) under the European Union’s Horizon 2020 research and innovation program under Grant Agreement ERC-StG-CENNS 803079, and the LabEx Lyon Institute of Origins (ANR-10-LABX-0066) of the Université de Lyon.


\end{document}